\documentclass[pdflatex,sn-mathphys]{sn-jnl}

\usepackage{mathtools}
\usepackage{hyperref}
\usepackage{rotating}
\usepackage{amsmath}
\usepackage{amssymb,float}
\usepackage[utf8]{inputenc}

\DeclareMathOperator\arctanh{arctanh}

\jyear{2022}%
\begin{document}

\title[Cosmological piecewise functions to treat the local Hubble tension]{Cosmological piecewise functions to treat the local Hubble tension}


\author*[1]{\fnm{Rodrigo} \sur{Sandoval-Orozco}}\email{rodrigo$\_$sandoval18@ciencias.unam.mx}

\author*[1]{\fnm{Celia} \sur{Escamilla-Rivera}}\email{celia.escamilla@nucleares.unam.mx}

\affil[1]{\orgdiv{Instituto de Ciencias Nucleares}, \orgname{Universidad Nacional
	Aut\'onoma de M\'exico}, \orgaddress{\street{Circuito Exterior C.U.}, \postcode{04510}, \state{M\'exico D.F.}, \country{M\'exico}}}


\abstract{The current cosmic time evolution of the Universe is described by the General Theory of Relativity when a cosmological principle is considered under a flat space time landscape. The set of   known as \emph{Friedmann equations}, contain the principles that lead to the construction of the standard $\Lambda$CDM model. However, 
 the current state-of-art regarding these  equations, even if it is a fundamental method, is based in solving analytically the differential equations by considering several forms of matter/energy components 
 or evaluating them in specific cosmic times where two or more components contribute at 
 the same rate. This latter can be carry out through  the approach of piecewice solutions, whose reduce the numerical integrals. In this 
 paper we discuss new solutions through special analytical functions  and constraint them with an updated compilation of observational Hubble observations in order to deal with the local $H_0$ tension reported.
}



\maketitle


\section{Introduction}
\label{sec:introduction}

The consensus in the scientific community is growing towards more complex and complete models to 
describe the Universe with more accuracy and precision in comparison to the standard $\Lambda$CDM model. This model, which, on one hand is in good agreement with our current observations \cite{Planck:2018vyg} \cite{SupernovaSearchTeam:1998fmf}, suffers with
of issues that have grown even bigger after the recent tensions, e.g. regarding the $H_0$ \cite{DiValentino:2021izs} and $S_8$ \cite{2021S8Tension} tensions. Using the improvement on the observations, the differences in the measurements of such parameters have been noticed by the early measurements from Planck Collaboration \cite{Planck:2018vyg} and late observations with supernovae Type Ia \cite{2018Pantheon}, where a discrepancy between $4\sigma$ and $6\sigma$ \cite{Riess:2020fzl} is observed in the measures of $H_0$. In addition, the measurements of the rate of the growth of the matter structure have also problems  \cite{2021S8Tension} that lead us to $3\sigma$ in the uncertainties.
In this context, our approach to these problems 
is to improve the way to solve the evolution  
equations that describe the epochs of the universe. 

The current state of cosmology focuses in  study the overall dynamics of the system that is our universe using Friedmann equations. However, the solutions to these equations are still derived in a less than straightforward manner. It is common in the literature (see e.g. \cite{Dodelson:1282338,Mukhanov:991646,Liddle:1010476}) that the solutions are presented in a \textit{piecewise form} by joining the separated solutions obtained via individual components of the fluids alone. This solution is correct in epochs where the selected fluid is dominant over the others, e.g at early times, the radiation (or relativistic matter) dominates, while at late times the nonrelativistic matter (including dark matter) dominates the fluid density of the Universe. In that direction, Galanti and Roncadelli \cite{galanti_2021_precision} offer another kind of solution: proposing a form of solving the differential equations using two different components at the same cosmic time. By doing this, the necessity is harder in a mathematical way, but suffers from physical justifications on the energy conditions of such fluids. In this line of thought it is important to mention that piecewise form solutions have been develop as a tool to classify the dynamics of inhomogeneous spherically symmetric universes \cite{Bochicchio:2011wx}.
In order to correct such conditions and derive new cosmological solutions, we present in this paper a more general way by including density curvature $\Omega_k$ to this landscape. In this path, we obtain a possible solution using the Weierstrass elliptic function $\oldwp$-function\cite{Steiner}, which requires a strict mathematical construction but with physical motivations.

This paper is divided as follows: in Sec.\ref{sec:piecewise} we present the standard  piecewise solutions for universes with three kinds of geometries, including radiation and matter components. In Sec.\ref{sec:Weierstrass} we present our Weierstrass elliptic solutions for the same epoch of the Universe described, and also we derive a general $H(t)$ solution. In Sec.\ref{sec:data-analysis} we perform a data analysis using an updated compilation of observational Hubble observations in order to constraints all the $H(z)$ solutions obtained. Furthermore, we include a discussion of such results by comparing our $H_0$ fitted values with the measurements derived by Planck Collaboration and through observations from HST of Cepheids \cite{Riess:2021jrx}\footnote{We refer to this $H_0$ prior as Riess et al \cite{Riess:2021jrx}.}. 


\section{Piecewise cosmological solutions}
\label{sec:piecewise}

The Friedmann-Lemaître-Robertson-Walker (FLRW) metric is one of the exact solutions for Einstein's field equations. This metric has the characteristic of being homogeneous and isotropic, describing a universe with the following evolution equation

\begin{equation}
\left( \frac{\dot{a}}{a} \right) = \frac{8\pi G}{3c^2} \varepsilon(t) - \frac{kc^2}{R_0^2 a^2},
\label{eq:original}
\end{equation}
where essentially, $a(t)$ is the scale factor, $G$ is the gravitational constant and $c$ the speed of light. The dot indicate derivatives with respect of the cosmic time. In this equation, $R_0^2$ denotes the scale factor at current time $t_0$ and $k$ is the curvature constant.
The usual form of Friedmann equations take place when we substitute $\varepsilon(t)$ with their analytic expression, in this case that expression is given by
\begin{equation}
\varepsilon(t) = \frac{3c^2 H_0^2}{8\pi G} \left( \frac{\Omega_{R,0}}{a^4} + \frac{\Omega_{M,0}}{a^3} + \Omega_{\Lambda} \right).
\label{eq:energy}
\end{equation}
Using this expression in Eq.(\ref{eq:original}) for a flat universe $k=0$ we arrive to
\begin{equation}
\left( \frac{\dot{a}}{a}\right)^2 = H_0^2 \left( \frac{\Omega_{R,0}}{a^4} + \frac{\Omega_{M,0}}{a^3} + \Omega_{\Lambda} \right),
\label{eq:substituted}
\end{equation}
where, $\Omega_i$ denotes the critical densities for each $i$ components, e.g. $i=M$ standard matter and $i=R$ for radiation.
Using (\ref{eq:energy}) we can obtain the scale factor when radiation and matter components contribute at same rate given by 
\begin{equation}
a_{RM} = a(t_{RM}) = \frac{\Omega_{R,0}}{\Omega_{M,0}},
\end{equation}
where subindexes $RM$ denotes this equal rate. 
For standard matter, including dark matter,
and dark energy we have the expression for the equivalence denoted by the subindexes $M\Lambda$ as
\begin{equation}
a_{M\Lambda} = a(t_{M\Lambda}) = \left( \frac{\Omega_{M,0}}{\Omega_\Lambda }  \right)^{1/3}.
\end{equation}
Using separately the epochs of radiation, matter and dark energy, the usual way to proceed is to consider the solutions according to the time dependence for each epoch as follows:
\begin{equation}
a_{pw} (t) = \left\{ \begin{matrix}
K_R t^{1/2} & \quad & t \leq \tilde{t}_{RM}, \\
K_M t^{2/3} & \quad & \tilde{t}_{RM} \leq t \leq \tilde{t}_{M\Lambda}, \\
K_\Lambda \exp \left( \Omega_\Lambda H_0t \right) & \quad & t > \tilde{t}_{M\Lambda},
\end{matrix} 
\right.
\label{eq:piecewise}
\end{equation}
where 
\begin{eqnarray}
K_R &=& \left( \frac{\Omega_{R,0}}{\Omega_{\Lambda}} \right)^{1/4} \left[ H_0t_0 + \frac13 \Omega_\Lambda^{-1/2}\ln \left( \frac{\Omega_{M,0}}{\Omega_\Lambda} \right) \right]^{-1/2} H_0^{1/2}, \\ 
K_M &=& \left( \frac{\Omega_{M,0}}{\Omega_{\Lambda}} \right)^{1/3} \left[ H_0t_0 + \frac13 \Omega_\Lambda^{-2/3}\ln \left( \frac{\Omega_{M,0}}{\Omega_\Lambda} \right) \right]^{-2/3} H_0^{2/3},\\ 
K_\Lambda &=& \exp \left( -\Omega_\Lambda^{1/2} H_0t_0 \right).
\end{eqnarray}
These expressions are obtained if we guarantee the continuity of the function $a(t)$ at times $t_{RM} $ and $t_{M\Lambda}$, which indicate the equivalence times between the components. 

Additionally, rewriting Eq.(\ref{eq:substituted}) in terms of $H(t)$ we can obtain
\begin{equation}
H_0 t = \int\limits_{0}^{\tilde{a}} \left( \frac{\Omega_{R,0}}{a^2} + \frac{\Omega_{M,0}}{\tilde{a}} + \Omega_{\Lambda}\tilde{a}^2 \right)^{-\frac{1}{2}} \,d\tilde{a},
\label{eq:friedmancompleta}
\end{equation}
which allows to express the time $t$ as function of $a(t)$. This expression is useful to perform numerical integration over the cosmic time of interest. To obtain our numerical solution, we start by considering a \textit{vanilla} model which denotes different components of the universe. 

\begin{figure*}
\centering 
\includegraphics[scale=0.4]{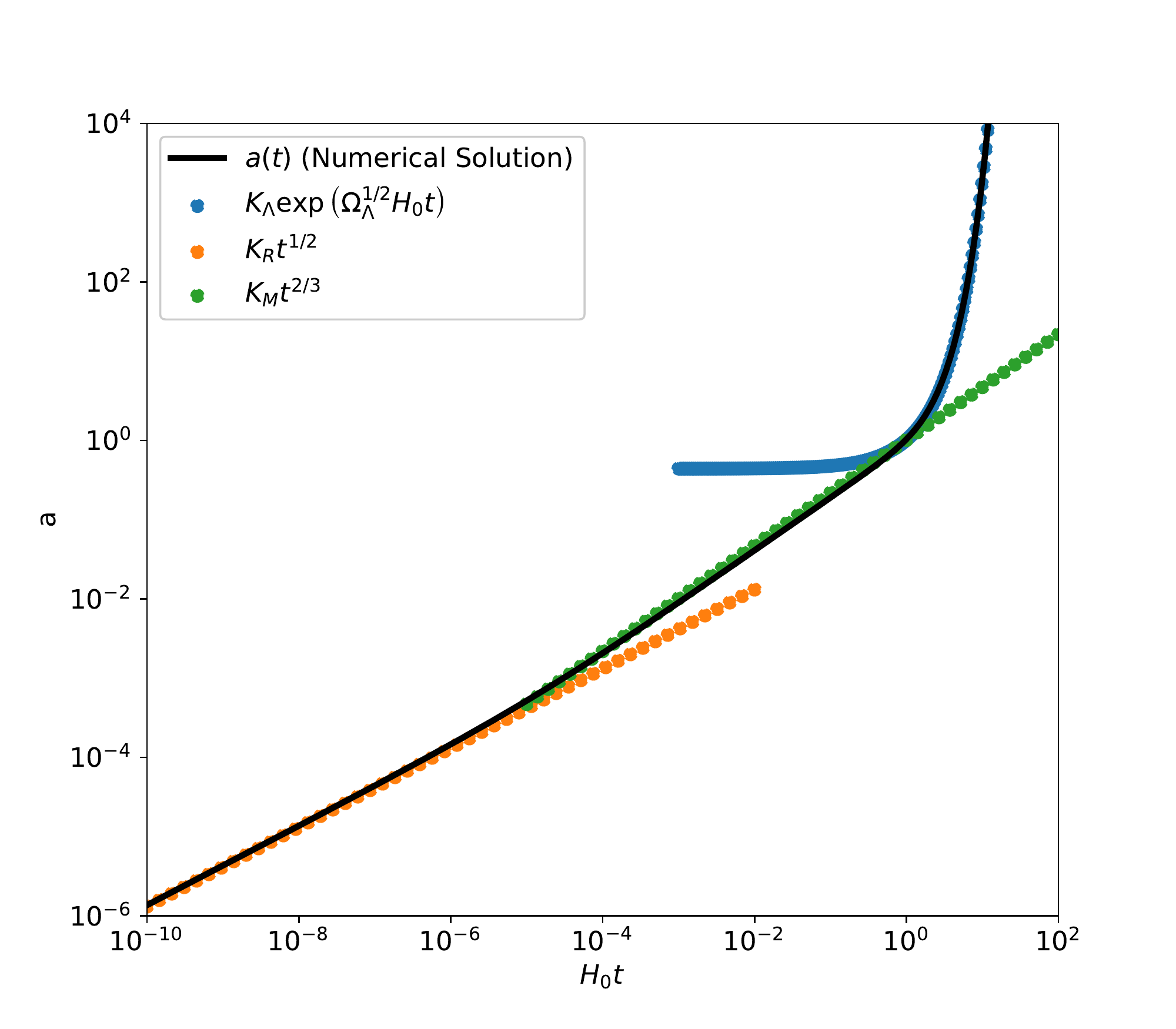}
\caption{Evolution of the piecewise solution compared to the numerical integration of (\ref{eq:friedmancompleta}) with Benchmark values \cite{galanti_2021_precision}. The orange curve is the first solution of (\ref{eq:piecewise}), where the radiation component dominates. The green curve represents the second solution where matter component dominates and finally, the blue curve represents the late solution where the universe is dominated by a cosmological constant.}
\label{fig:firstpiecewiese}
\end{figure*}

In \cite{galanti_2021_precision} was used the Benchmark model as a vanilla model. In order to compare our results, we will employ the same fitted values: $\Omega_{R,0} = 8.45 \times 10^{-5} $, $\Omega_{M,0} = 0.3 $ and $\Omega_\Lambda = 0.70 $. Also, we consider a set of realistic priors from Planck \cite{Planck:2018vyg}: $\Omega_\Lambda = 0.679 $, $\Omega_{M,0} = 0.321 $, $\Omega_{k} = -0.056 $ and $\Omega_{R,0} =1-\Omega_\Lambda-\Omega_{M,0}-\Omega_{k}$. The evolution of both analyses are given in Figure \ref{fig:firstpiecewiese}.

To study the evolution of (\ref{eq:friedmancompleta}) we can calculate the piecewise $H_{pw}(t)$ function using $a_{pw}(t)$ form given in (\ref{eq:piecewise}). After straightforward calculation we obtain
\begin{equation}
\dot{a}_{pw} (t) = \left\{ \begin{matrix}
\frac12 \left( \frac{\Omega_{R,0}}{\Omega_{\Lambda}} \right)^{1/4} \left[ H_0t_0 + \frac13 \Omega_\Lambda^{-1/2}\ln \left( \frac{\Omega_{M,0}}{\Omega_\Lambda} \right) \right]^{-1/2} H_0^{1/2} t^{-1/2} & \quad & t \leq \tilde{t}_{RM}, \\
\frac{2}{3} \left( \frac{\Omega_{M,0}}{\Omega_{\Lambda}} \right)^{1/3} \left[ H_0t_0 + \frac13 \Omega_\Lambda^{-2/3}\ln \left( \frac{\Omega_{M,0}}{\Omega_\Lambda} \right) \right]^{-2/3} H_0^{2/3} t^{-1/3} & \quad & \tilde{t}_{RM} \leq t \leq \tilde{t}_{M\Lambda}, \\
K_\Lambda \Omega_\Lambda H_0 \exp \left( \Omega_\Lambda H_0t \right) & \quad & t > \tilde{t}_{M\Lambda},
\end{matrix} 
\right.
\label{eq:adotpiecewise}
\end{equation}
and therefore, we can express the Hubble parameter piecewise as
\begin{equation}
H_{pw} (t) = \left\{ \begin{matrix}
\frac{1}{2t} & \quad & t \leq \tilde{t}_{RM}, \\
 \frac{2}{3t} & \quad & \tilde{t}_{RM} \leq t \leq \tilde{t}_{M\Lambda}, \\
\Omega_\Lambda H_0 & \quad & t > \tilde{t}_{M\Lambda}.
\end{matrix} 
\right.
\label{eq:Htpiecewise}
\end{equation}
Notice that we obtained a form of piecewise solution for Hubble parameter, which can be measure. We are going to discuss this aspect in the next two models.


\subsection{Model I: Without curvature}
\label{subsec:model1}

\begin{figure*}
\centering 
\includegraphics[scale=0.4]{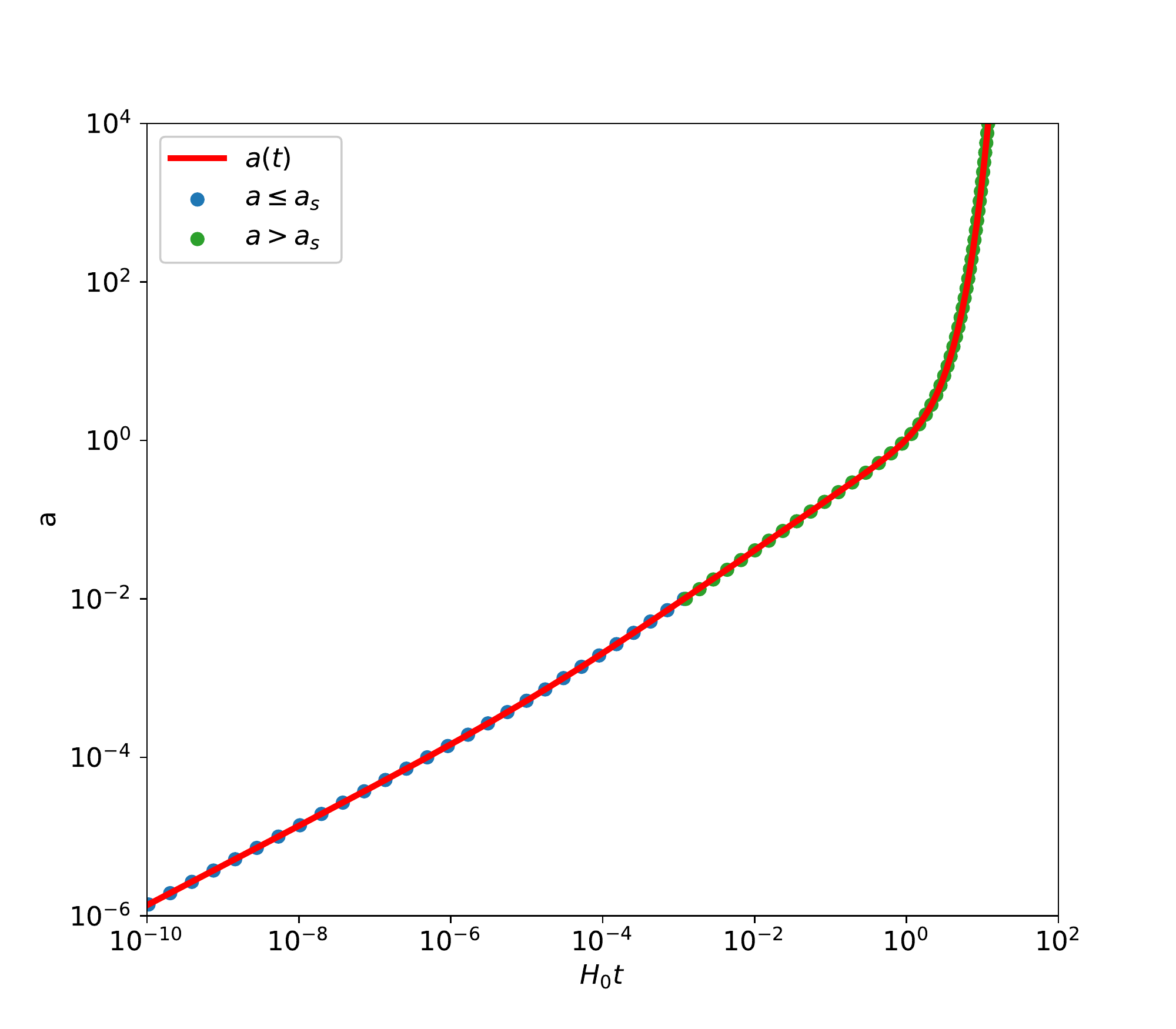}
\caption{Behaviour of the solution obtained from Eq.(\ref{eq:intermediatesolution}) and the numerical integration. Notice that both are indistinguishable.}
\label{fig:intermediateplot}
\end{figure*}

In \cite{galanti_2021_precision} it was described the process to obtain a solution mixing the matter components of the universe. In comparison to this way to proceed, in here we propose to deal with our piecewise solutions by integrating them in two parts: \textit{(i)} up to the matter-radiation dominated universe, and \textit{(ii)} up to the matter-dark energy dominated universe; proposing  a time $t_s $ between $t_{RM} $ and $t_{M\Lambda} $. Also, we divided the epochs so $t \leq t_s $ implies matter/radiation domination and, for $t > t_s $ solely matter/dark energy domination: 
\begin{equation}
H_0 t = \left\{ \begin{matrix} 
\displaystyle\int_{0}^{a(t)} d\tilde{a}\left( \frac{\Omega_{R,0}}{\tilde{a}^2} + \frac{\Omega_{M,0}}{\tilde{a}} \right)^{-1/2} \quad a \leq a_s,   \\
H_0t_s + \displaystyle\int_{a_s}^{a(t)} d\tilde{a}\left( \frac{\Omega_{M,0}}{\tilde{a}} + \Omega_\Lambda \tilde{a}^2 \right)^{-1/2} \quad a > a_s,
\end{matrix} \right.
\end{equation}
These integrals yield the following expressions: 
\begin{equation}
H_0 t = \left\{ \begin{matrix} 
\frac{2}{3 \Omega_{M,0}^2} \left[ \left( \Omega_{M,0}a - 2\Omega_{R,0} \right)\left(\Omega_{M,0}a + \Omega_{R,0} \right)^{1/2} + 2\Omega_{R,0}^{3/2} \right], \\
H_0 t_s +  \frac{2}{3\Omega_\Lambda^{1/2}} \ln{\left[\frac{\Omega_\Lambda a^{3/2}+ \Omega_\Lambda^{1/2} \left(\Omega_{M,0} + \Omega_\Lambda a^3 \right) }{\Omega_\Lambda a_s^{3/2}+ \Omega_\Lambda^{1/2} \left(\Omega_{M,0} + \Omega_\Lambda a_s^3 \right)^{1/2}}  \right]}.
\end{matrix} \right.
\label{eq:intermediatesolution}
\end{equation}
The goal is to express them as a function of $a(t)$. Here, we can introduce an intermediate step, to see that the integral in question is calculated in the right way. We can plot this solution and compare with the numerical integration. This result is presented in Figure \ref{fig:intermediateplot}. As it is shown, the numerical integration and the split up analytical solution are consistent with each other. 

According to the latter, the $a(t)$ functions for this case are given by
\begin{equation}
a(t) = \left\{ 
\begin{matrix} 
\frac{\Omega_{R,0}}{\Omega_{M,0}} \left\{ 1 - 2\sin{\left[\frac13\arcsin{\left(X(t)\right)} \right]} \right\} \quad 0 \leq t \leq t_*, \\
\frac{\Omega_{R,0}}{\Omega_{M,0}} \left\{ 1 + 2\cos{\left[\frac13\arccos{\left(X(t)\right)} \right]} \right\} \quad t_* < t \leq t_s, \\
\left\{a_s^{3/2} \cosh{\left[\frac32 \Omega_\Lambda^{1/2} H_0(t-t_s)\right]} + \left(a_s^3 + \frac{\Omega_{M,0}}{\Omega_\Lambda} \right)^{1/2} \sinh{\left[\frac32 \Omega_\Lambda^{1/2} H_0(t-t_s)  \right]}  \right\}^{2/3}, \\ t > t_s,
\end{matrix} 
\right.
\label{eq:fullexactsolution}
\end{equation}
where $X(t) = 1 -3\frac{\Omega_{M,0}^2}{\Omega_{R,0}^{3/2}}H_0t + \frac{9}{8}\frac{\Omega_{M,0}^4}{\Omega_{R,0}^3}H_0^2 t^2$.

The next step is to express $a(t)$ to obtain a complete analytical solution to the system. This process described in \cite{galanti_2021_precision} is not trivial due that it requires different algebra and calculus steps to reach the full exact solution that is shown in Eq.(\ref{eq:fullexactsolution}).  At the time $t_* = 4\Omega_{R,0}^{3/2}/(3\Omega_{M,0}^2 H_0)$ we can obtain a function of the scale factor whose dependence guarantee the continuity of the solution in all the desired observational interval. 

Using the relation $H = \dot{a}/a$ we can explore the cosmological properties of the obtained solutions. First, we can obtain the first derivatives $\dot{a}(t)$ as: 
\begin{equation}
\dot{a}(t) = \left\{ \begin{matrix} 
-\frac{2 \Omega_{R,0}}{3 \Omega_{M,0}} \left( \frac{\cos\left[ \frac{1}{3}\arcsin(X(t))  \right] \dot{X}(t)}{\sqrt{1 - X^2(t)}} \right) \quad 0 \leq t \leq t_*, \\
\frac{2 \Omega_{R,0}}{3 \Omega_{M,0}} \left( \frac{ \sin\left[ \frac{1}{3}\arccos(X(t))  \right] \dot{X}(t)}{\sqrt{1 - X^2(t)}} \right) \quad t_* < t \leq t_s, \\
\frac{\Omega_{\Lambda}^{1/2}H_0 \left(a_s^{3/2} \sinh{\left[\frac32 \Omega_\Lambda^{1/2} H_0(t-t_s) \right]} +\left(a_s^3 + \frac{\Omega_{M,0}}{\Omega_\Lambda} \right)^{1/2} \cosh{\left[\frac32 \Omega_\Lambda^{1/2} H_0(t-t_s) \right]} \right)}{\left( a_s^{3/2} \cosh{\left[\frac32 \Omega_\Lambda^{1/2} H_0(t-t_s)\right]} + \left(a_s^3 + \frac{\Omega_{M,0}}{\Omega_\Lambda} \right)^{1/2} \sinh{\left[\frac32 \Omega_\Lambda^{1/2} H_0(t-t_s)  \right]}  \right)^{1/3}} \quad t > t_s,
\end{matrix} \right.
\label{eq:hubble_analytic}
\end{equation}
where $\dot{X}(t) = -3\frac{\Omega_{M,0}^2}{\Omega_{R,0}^{3/2}}H_0 + \frac{9}{4}\frac{\Omega_{M,0}^4}{\Omega_{R,0}^3}H_0^2 t$.
So, the Hubble factor can be seen as
\begin{equation}
H(t) = \left\{ \begin{matrix} 
\frac{2 \dot{X}(t)}{3 \sqrt{1 -X^2(t)}} \frac{\sin\left[ \frac{1}{3}\arccos(X(t))  \right]}{1 - 2\sin{\left[\frac13\arcsin{\left(X(t)\right)} \right]}}  \quad 0 \leq t \leq t_*, \\
\frac{2 \dot{X}(t)}{3 \sqrt{1 -X^2(t)}} \frac{\cos\left[ \frac{1}{3}\arcsin(X(t))  \right]}{1 + 2\cos{\left[\frac13\arccos{\left(X(t)\right)} \right]}} \quad t_* < t \leq t_s, \\
\Omega_{\Lambda}^{1/2}H_0 \frac{a_s^{3/2} \sinh{\left[\frac32 \Omega_\Lambda^{1/2} H_0(t-t_s) \right]} +\left(a_s^3 + \frac{\Omega_{M,0}}{\Omega_\Lambda} \right)^{1/2} \cosh{\left[\frac32 \Omega_\Lambda^{1/2} H_0(t-t_s) \right]}}{a_s^{3/2} \cosh{\left[\frac32 \Omega_\Lambda^{1/2} H_0(t-t_s)\right]} + \left(a_s^3 + \frac{\Omega_{M,0}}{\Omega_\Lambda} \right)^{1/2} \sinh{\left[\frac32 \Omega_\Lambda^{1/2} H_0(t-t_s)  \right]}} \quad t > t_s.
\end{matrix} \right.
\label{eq:hubble_analytic}
\end{equation}

\begin{figure*}
\centering 
\includegraphics[scale=0.4]{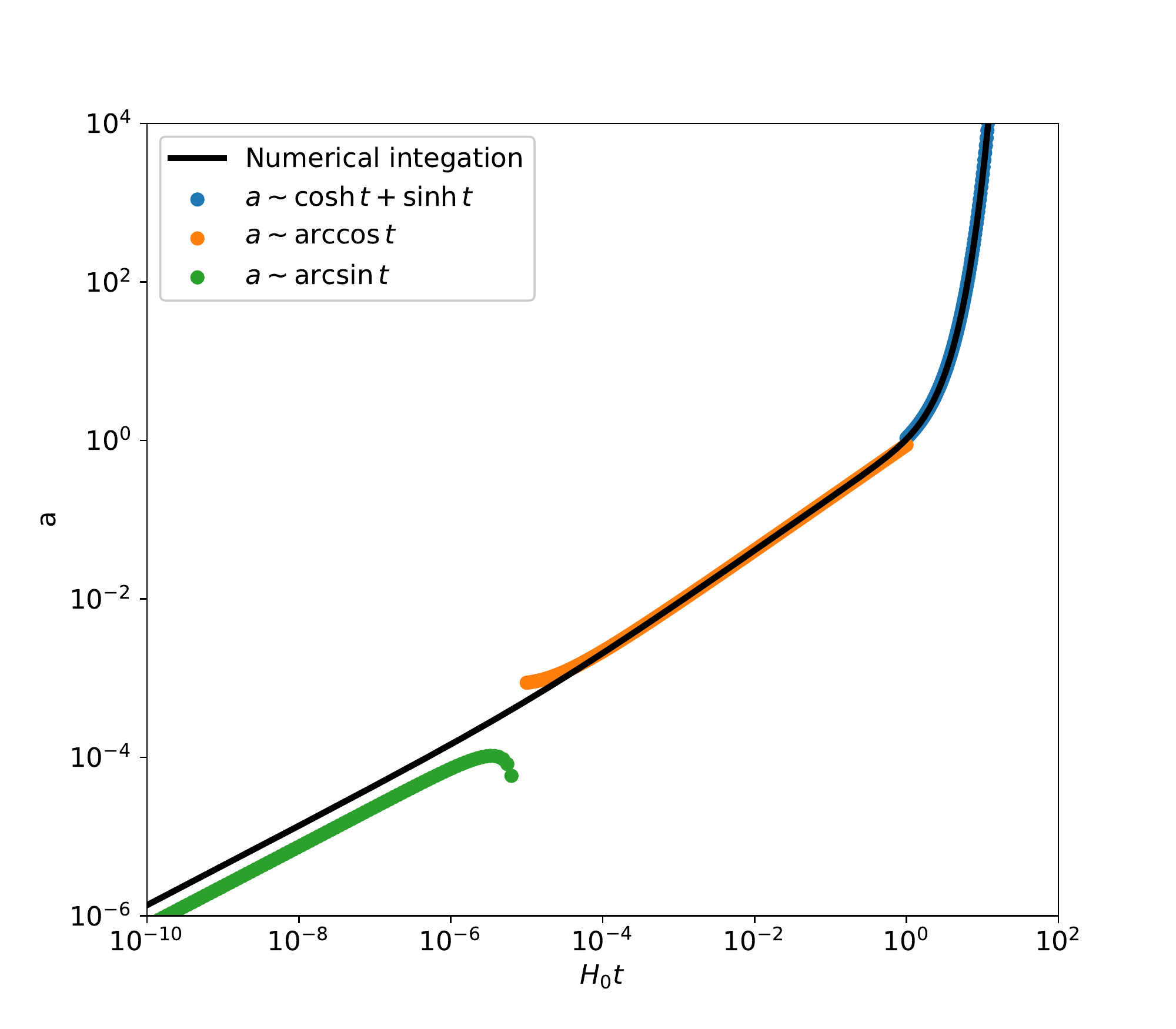}
\caption{Behaviour of the full analytic solutions and the numerical integration of (\ref{eq:fullexactsolution}). The green, blue and orange curves are the solutions presented in Eq.(\ref{eq:fullexactsolution}) for radiation-matter mixed dominated, the matter-radiation late solution and the late solution mixing matter and cosmological constant, respectively.}
\label{fig:problem1}
\end{figure*}


\subsection{Model II: With curvature}
\label{subsec:model2}

In follow we present a generalization of the calculations already developed. 
As we notice in the previous subsection, the first approximation to solve the Friedmann equation analytically is to split the integral and assume a flat universe. The approach next consider, generalise this aspect. Eq. (\ref{eq:friedmancompleta}) can be expressed with a new density for the curvature as
\begin{equation}
H_0 t = \int\limits_{0}^{\tilde{a}} \left( \frac{\Omega_{R,0}}{\tilde{a}^2} + \frac{\Omega_{M,0}}{\tilde{a}} + \Omega_k + \Omega_{\Lambda}\tilde{a}^2 \right)^{-\frac{1}{2}} \,d\tilde{a}.
\label{eq:friedmanncurvature}
\end{equation}
A way to solve this equation is to split up the integral in a matter-radiation and matter-dark energy dominated universe
\begin{eqnarray}
H_0 t = \left\{ \begin{matrix} 
\displaystyle\int_{0}^{a(t)} d\tilde{a}\left( \frac{\Omega_{R,0}}{\tilde{a}^2} + \frac{\Omega_{M,0}}{\tilde{a}} + \Omega_k \right)^{-1/2} \quad a \leq a_{RM},   \\

H_0t_s + \displaystyle\int_{a_s}^{a(t)} d\tilde{a}\left( \frac{\Omega_{M,0}}{\tilde{a}} + \Omega_\Lambda \tilde{a}^2 + \Omega_k \right)^{-1/2} \quad a > a_s.
\end{matrix} \right.
\label{eq:splittedcurvature}
\end{eqnarray}
We assume this by using the evolution of the critical density parameters to see how they evolve and to study how different components dominate in different epochs. 

\begin{figure}[H]
    \centering
    \includegraphics[scale=0.4]{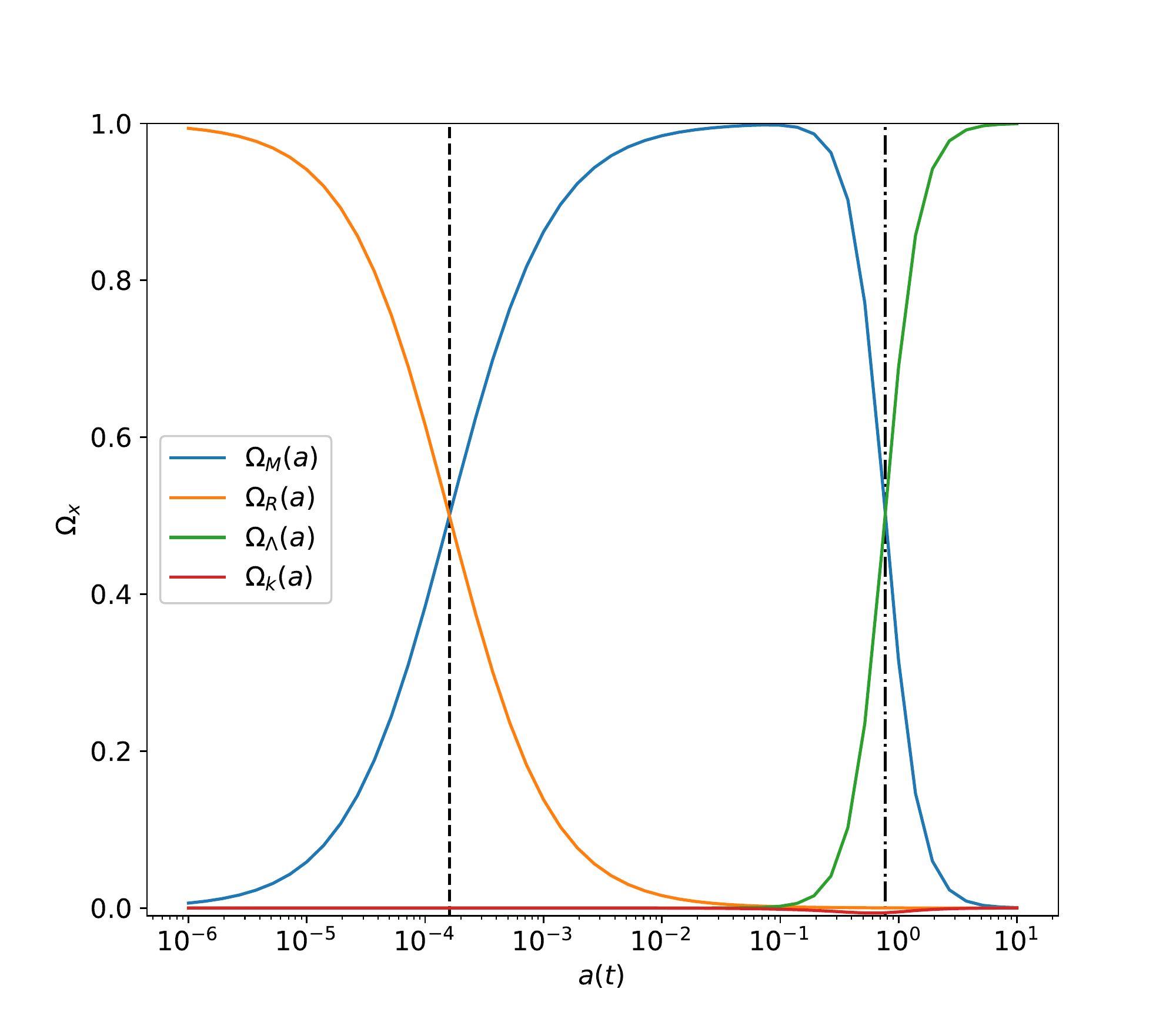}
    \caption{Evolution of the critical density parameters in terms of $a$. The dashed and dashed-dotted lines are radiation-matter equivalence time and matter-dark energy equivalence time, respectively. Here, the blue color line denotes the critical density matter parameter, the orange color line denotes the critical density radiation parameter and the green and red colors are cosmological constant and curvature, respectively.}
    \label{fig:evolution}
\end{figure}

Figure \ref{fig:evolution} confirms what we already described: the universe has an epoch of radiation domination, afterwards a matter domination epoch and latter on describes a domination of dark energy. This is the reason why we can split the integrals as a piecewise solution. The first integration can be done with common calculus techniques. The result is given by  
\begin{eqnarray}
H_0t &=& \frac{2\sqrt{\lvert\Omega_k\rvert}\left(a^2 \Omega_k + a\Omega_{M,0} + \Omega_{R,0} \right) }{2a \lvert\Omega_k\rvert^{3/2} \sqrt{\frac{a\Omega_{M,0} +\Omega_{R,0}}{a^2} + \Omega_k } } \nonumber\\
&& - \frac{\Omega_{M,0} \sqrt{a^2 \Omega_k + a\Omega_{M,0} + \Omega_{R,0}} \arctanh{\left( \frac{2a\Omega_k + \Omega_{M,0}}{2\sqrt{\lvert\Omega_k\rvert}\sqrt{\Omega_{R,0} + a(\Omega_{M,0} + \Omega_k a) } } \right)} }{2a \lvert\Omega_k\rvert^{3/2} \sqrt{\frac{a\Omega_{M,0} +\Omega_{R,0}}{a^2} + \Omega_k } } . \quad\quad
\label{eq:solution1}
\end{eqnarray}
This equation can be transform into an easier version by transforming the trigonometric expression to an exponential one\footnote{This step is straightforward since we only used the identity $2 \text{arctanh}{z} = \ln{(1+z)} - \ln{(1-z)}$. } which can be expressed also in an easier form as:
\begin{eqnarray}
 H_0t &=& \frac{\sqrt{a^2\lvert\Omega_k\rvert + a\Omega_{M,0} + \Omega_{R,0}}}{\lvert\Omega_k\rvert} - 
 \nonumber \\ &&
 \frac{\Omega_{M,0}}{2\lvert\Omega_k\rvert^{3/2}} \ln\left(2\sqrt{\frac{\lvert\Omega_k\rvert\left(a^2\lvert\Omega_k\rvert + a\Omega_{M,0} + \Omega_{R,0} \right) }{4\lvert\Omega_k\rvert\Omega_{R,0} - \Omega_{M,0}}} + \frac{2a\lvert\Omega_k\rvert + \Omega_{M,0}}{\sqrt{4\Omega_{R,0}\rvert\Omega_k\lvert - \Omega_{M,0}^2}}\right). \quad\quad
\label{eq:soleasy}
\end{eqnarray}

\begin{figure}[H]
    \centering
    \includegraphics[scale=0.4]{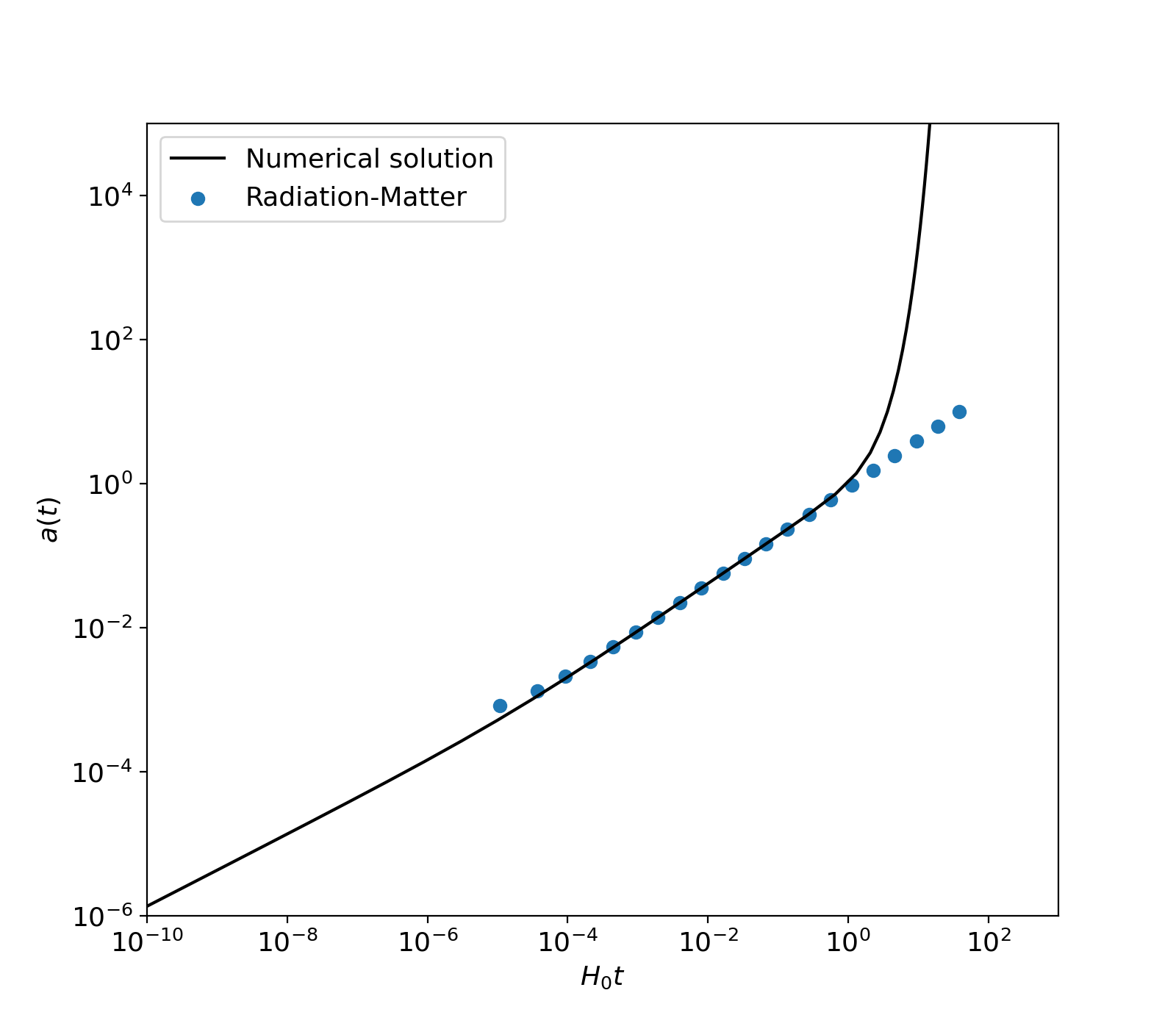}
    \caption{Numerical integration of Eq.(\ref{eq:soleasy}). This is an intermediate step to study the behaviour of the solution using piecewise conditions.}
    \label{fig:earlyproblem}
\end{figure}

As we can notice in Figure \ref{fig:earlyproblem} there is a problem regarding the integral at early times with that amount of components. This issue comes due the analytical reversal, since is not possible to obtain an explicit $a(t)$ solution in this interval, at least analytical. 

The second part of this integral represents a problem, because it does not have an analytic solution. The complete integral only have numerical solution by reducing it to an elliptic integral of the second kind. This kind of integral cannot be solved analytically. The only way to obtain a solution for the second $H_0 t$ expression is to separate the integral in a matter dominated epoch and a dark energy dominated epoch. For this purpose we have
 \begin{equation}
H_0 t = \left\{ \begin{matrix} H_0 t_{RM}+ \displaystyle\int_{a_{RM}}^{a(t)} d\tilde{a} \left( \frac{\Omega_{M,0}}{\tilde{a}} + \Omega_k \right)^{-1/2}, &   &a_0 \geq a \geq a_{RM}, \\
H_0t_{0} + \displaystyle\int_{a_0}^{a(t)} d\tilde{a} \left( \Omega_\Lambda \tilde{a}^2  + \Omega_k \right)^{-1/2}, &   & a \geq a_0, 
\end{matrix} \right.
\label{eq:splitsecondpart}
 \end{equation}
where $t_0 $ is the time where dark energy and matter are contributing at the same rate. The solution to these integrals is possible to obtain with common techniques. Following
\begin{equation}
H_0 t = \left\{ \begin{matrix} 
H_0 t_{RM} + \frac{a\sqrt{\Omega_k + \frac{\Omega_{M,0}}{a}} - a_{RM} \sqrt{\Omega_k + \frac{\Omega_{M,0}}{a_{RM}}}}{\Omega_k}+ & & \\ \frac{\Omega_{M,0}}{\lvert\Omega_k\rvert^{3/2}} \left[ \arctanh{\left(\sqrt{\frac{\Omega_k +\frac{\Omega_M}{a} }{\lvert\Omega_k\rvert}}\right)} - \arctanh{\left(\sqrt{\frac{\Omega_k +\frac{\Omega_M}{a_{RM}} }{\lvert\Omega_k\rvert}}\right)}  \right], &   & a_{RM} < a \leq a_0, \\
H_0t_{0} + \frac{1}{\Omega_\Lambda^{1/2}} \ln{\left( \frac{\sqrt{\frac{\Omega_\Lambda}{\lvert\Omega_k\rvert}a^2 + 1} + \sqrt{\frac{\Omega_\Lambda}{\lvert\Omega_k\rvert}} a}{\sqrt{\frac{\Omega_\Lambda}{\lvert\Omega_k\rvert}a_0^2 + 1} + \sqrt{\frac{\Omega_\Lambda}{\lvert\Omega_k\rvert}} a_0}  \right)}, &   & a > a_0. 
\end{matrix} \right.
\label{eq:solution2}
\end{equation}

\begin{figure}[H]
\centering
\includegraphics[scale=0.4]{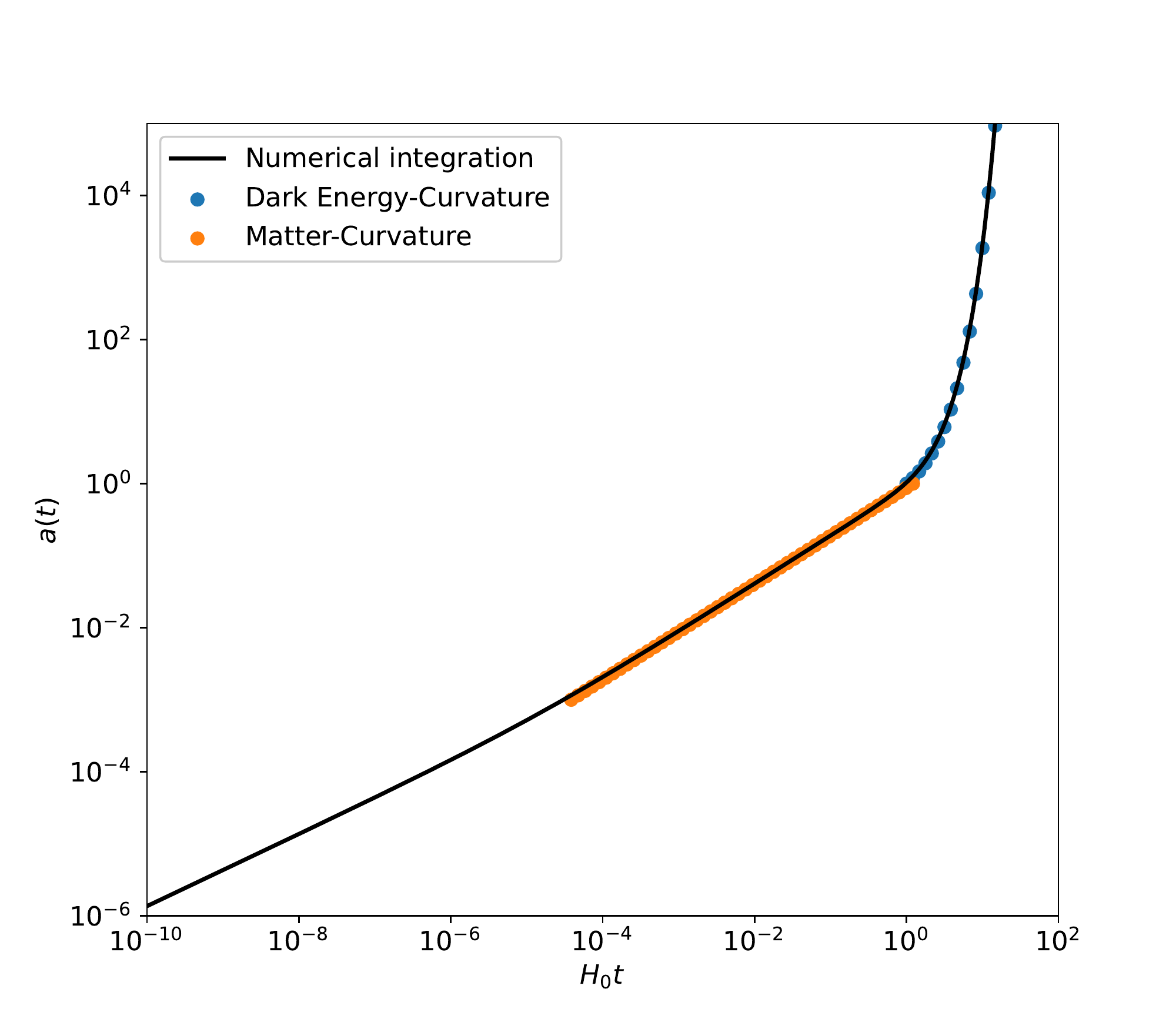}
\caption{Comparison between numerical integration and the results of integrals for latter epochs of the universe. This process is plotted using Planck values \cite{Planck:2018vyg}. The blue dots represent the late solution using curvature and a cosmological constant, while the orange dots represent the solution with matter and curvature.}
\label{fig:step1resultsofintegrals}
\end{figure}

Before expressing the analytical solutions, we can perform an intermediate step to compute the results of the different integrals. The comparison between the numerical integration is shown in Figure \ref{fig:step1resultsofintegrals}. 

Performing some algebra we can reach a direct analytic expression for $a$ with direct dependence on $t$ from Eq.(\ref{eq:solution2}) as
\begin{equation}
a(t) = \frac{e^{-F-G}\left(-1+e^{2F+2G} \right)}{2\sqrt{\frac{\Omega_\Lambda}{\lvert\Omega_k\rvert}}} = \frac{\sinh{\left(F + G\right)}}{\sqrt{\frac{\Omega_\Lambda}{\lvert\Omega_k\rvert}}},
\end{equation}
where $G = \ln{\left(\sqrt{\frac{\Omega_\Lambda}{\lvert\Omega_k\rvert}a_0^2 + 1} + \sqrt{\frac{\Omega_\Lambda}{\lvert\Omega_k\rvert}a_0}\right)}$ and $F =\sqrt{ \Omega_\Lambda} H_0(t-t_0)$. This is one of the general solutions to Friedmann equation with curvature. 
In order to build an explicit function for $a(t)$, we need to consider some approximations for the expression with arctanh function up to first order and around $a_{RM}$ in Eq. (\ref{eq:solution2}) 
\begin{equation}
    \arctanh{\sqrt{\frac{\Omega_k + \frac{\Omega_{M,0}}{a}}{\Omega_k}}} \cong \arctanh{\sqrt{\frac{\Omega_k + \frac{\Omega_{M,0}}{a_{RM}}}{\Omega_k}}} + \frac{\sqrt{\Omega_k}(a-a_{RM})\sqrt{\Omega_k + \frac{\Omega_{M,0}}{a_{RM}}}}{2\left(\Omega_k a_{RM} +\Omega_{M,0}\right)}+ \hdots
\end{equation}
and using this, we obtain 
\begin{equation}
    D(t) = a\sqrt{\Omega_k + \frac{\Omega_{M,0}}{a}} + a\frac{\Omega_{M,0}}{2(\Omega_k a_{RM} +\Omega_{M,0} )} \sqrt{\Omega_k + \frac{\Omega_{M,0}}{a_{RM}}},
\end{equation}
by using 
\begin{equation}
D(t) =  H_0(t-t_{RM}) + \left( 1+ \frac{\Omega_{M,0}}{2(\Omega_k a_{RM} +\Omega_{M,0})} \right)a_{RM} \sqrt{\Omega_k + \frac{\Omega_{M,0}}{a_{RM}}}.
\end{equation}
The third order polynomial has three solutions: two complex and one real. By considering 
\begin{equation}
    \beta := \frac{\Omega_{M,0}}{2(\Omega_k a_{RM}+\Omega_{M,0})} \sqrt{\Omega_k + \frac{\Omega_{M,0}}{a_{RM}}},
\end{equation}
we obtain an easy relation of second grade for $a(t)$ as
\begin{equation}
\begin{split}
    \left(D(t) -\beta a \right)^2 = a^2 \Omega_k + a\Omega_{M,0} \quad \Rightarrow \\ \quad \left(\beta^2 -\Omega_k \right)a^2 - (2\beta D(t) + \Omega_{M,0})a + D^2(t) = 0.
\end{split}
\end{equation}
This solution has one physically possible solution: $a(t) > 0$ for $t > 0$ given by
\begin{equation}
    a(t) = \frac{\sqrt{4D^2(t)(\Omega_k - \beta) +(2\beta D^2(t)+\Omega_{M,0})^2}-2\beta D(t) - \Omega_{M,0}}{2(\Omega_k - \beta)}.
\label{eq:matter-curvaturesolution}
\end{equation}
We have already two analytic expressions for the scale factor in terms of the cosmic time. Again, for this solution we obtain an expression for the Hubble parameter
\begin{equation}
    \dot{a}(t) = \frac{\left(\beta + \frac{\Omega_k + (\Omega_{M,0}-1)\beta + 2\beta^2 D(t)}{\sqrt{4(\Omega_k-\beta)D(t) + (\Omega_{M,0} + 2\beta D(t))^2}} 
    \right)H_0}{\Omega_k - \beta}.
\end{equation}
For the complete analytic solutions we can obtain an expression for $H(t)$ to explore the behaviour for the Hubble parameter
\begin{equation}
    H(t) = \frac{\dot{a}}{a} =\left\{ \begin{matrix} 
    \frac{H_0\left(\sqrt{4(\Omega_k -\beta)D(t) + (\Omega_{M,0}+2\beta D(t))^2} +2\beta D(t) \right)}{2D(t)\sqrt{4(\Omega_k - \beta)D(t) + (\Omega_{M,0} +2\beta D^2(t))}}, \quad a_{RM} < a < a_0, \\
    \sqrt{\Omega_\Lambda}H_0\coth{\left(F+G\right)}, \quad a > a_0.
    \end{matrix} \right.
\end{equation}
So, for our attempt of piecewise solution, we obtain finally an analytical expression for the Hubble parameter.

\section{Weierstrass-like solutions}
\label{sec:Weierstrass}

\begin{figure}[H]
    \centering
    \includegraphics[scale=0.4]{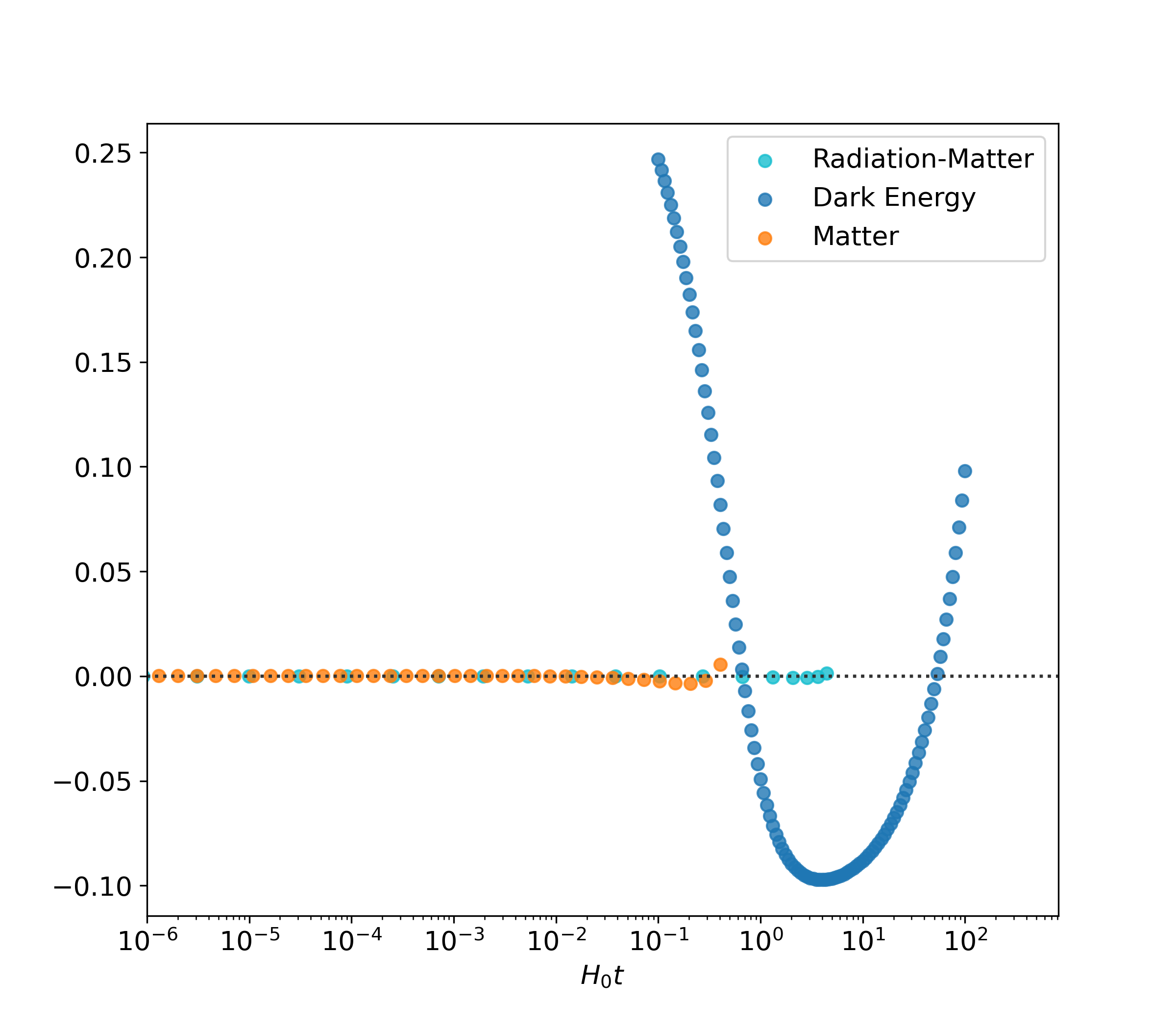}
    \caption{Residual values between different parts of the piecewise solutions analysed. The orange and light blue color dots denote the early universe, dominated by radiation and matter, meanwhile the dark blue color dots denotes the late solution.}
    \label{fig:residual}
\end{figure}

In Figure \ref{fig:problem1}, we can notice that at early times we have problems on the analytical solution. This is expected since the integration of Friedmann equations (\ref{eq:friedmancompleta}) and (\ref{eq:friedmanncurvature}) are not well-defined at $a \to 0$. However, a useful way to deal with this issue is by considering our piecewise technique. 
To begin, we compute the residual between analytical and numerical solutions in Figure \ref{fig:residual}. Here, the residual for the late solution is larger than the numerical integration. Notice that the matter solution exhibits a better behaviour, but the early time universe starts to present problems because the analytical solution does not work for the most part when combining matter, radiation and curvature. 

Strictly speaking, Friedmann equations are continuous in the interval $(0,\infty]$ and the set of points in which these are discontinuous
fulfils Lebesgue's integrability. The problem is more in the sense of finding the analytic solution instead of using a  non integrability method. Therefore, the function is integrable in the range $(0,\infty)$, in such a way that we can
obtain a complete solution using the Weierstrass  elliptic $\oldwp$-function\cite{Steiner}.  Steiner's approach is more general at least, by first declaring that Friedmann equations looks like
\begin{equation}
    H_0t = \int_{0}^{a(t)} \frac{da}{\sum_{j=0}^4 \Omega_j a_0^j a^{4-j} a}.
\end{equation}
This means that the integral can be written in a easier way using $t = F(a(t)$, where $F(z)$ is an elliptic integral. Finding $a(t)$ in this case is equivalent to compute $a(t) = F^{-1}(t)$, which is an elliptic function with
\begin{equation}
    a(t) = \frac{A}{2} \frac{\oldwp(t) - E\oldwp'(t) + \left(ABE^2 + \Omega_k/12 - D/2 \right) }{\left(\oldwp(t) + \Omega_k/12 - D/2\right)^2 - C},
\end{equation}
where $\oldwp(t) = \oldwp(t;g_2,g_3)$ in the interval $[0,\infty]$ for $H_0t$. Also, $g_2,g_3$ are invariants defined as
\begin{eqnarray}
    g_2 &:=& \frac{\Omega_k^2}{12} + 4C - 2AB + D(3D-\Omega_k), \\
    g_3 &:=& \frac{\Omega_k}{216} - \frac{8\Omega_kC + A^2\Omega_\Lambda}{12} - \frac{\Omega_k AB}{6} -A^2B^2E^2  \nonumber \\ && 
    +D\left( AB - \frac{\Omega_k^2}{12} + \frac{\Omega_k D}{2} - D^2 + 4C\right),
\end{eqnarray}
whose constants can be written in terms of the cosmological parameters:
\begin{eqnarray}\label{eq:notations}
A &=& \frac{1}{2}\Omega_{M,0} a_0^3, \\
B &=& \frac{1}{4} \Omega_{\phi}a_0, \\
C &=& \frac{1}{12} A^2 E^2 \Omega_\Lambda, \\
D &=& \frac{1}{6} \Omega_s a_0^2, \\
E &=& 2\frac{\sqrt{\Omega_{R,0}}}{\Omega_{M,0}} a_0^{-1}. \label{eq:notations2}
\end{eqnarray}
If we neglect $\Omega_s$ and $\Omega_\phi$ we can express the complete analytic solution to Friedmann equations as: 
\begin{equation}
    a(t) = \frac{A}{2} \frac{\oldwp(t) - E\oldwp'(t) + \left(\frac{\Omega_k}{12} \right) }{\left(\oldwp(t) + \frac{\Omega_k}{12} \right)^2 - C},
\label{eq:completeanalyticalsolution}
\end{equation}
where $g_2$ and $g_3$ are also modified and $\oldwp(\eta) = \oldwp(\eta;g_2,g_3)$ is given by
\begin{equation}
    \oldwp(z;\omega_1,\omega_2) = \frac{1}{z^2} + \sum_{m,n \neq 0}^{\infty} \left[\frac{1}{\left(z -2m\omega_1 -2n\omega_2 \right)^2} - \frac{1}{\left(2m\omega_1 +2n\omega_2 \right)^2} \right].
\end{equation}
For simplicity, it is easier to take the series expansion of $\oldwp(\eta)$ by considering
\begin{equation}
    \oldwp(z) = \frac{1}{z^2} + \sum_{k=2}^{\infty} c_k z^{2k-2},
\end{equation}
where $c_2 = \frac{\omega_2}{20}$ and $c_3 = \frac{\omega_3}{28}$ and
\begin{equation}
    c_k = \frac{3}{(2k+1)(k-3)} \sum_{m=2}^{k-2} c_m c_{k-m}.
\end{equation}
We can use as many $c_k$ in order to achieve a desired precision. However, to study the cosmological properties of these solutions, we can start by calculating $\dot{a}(t)$ and $H(t)$. 
By the properties of $\oldwp(z)$ we can write
\begin{equation}
    \oldwp(w) = \frac{1}{w^2} + \frac{g_2}{20}w^2 + \frac{g_3}{28}w^4 + \hdots, \quad w \in \mathbb{C},
\end{equation}
with the first derivative as
\begin{equation}
    \oldwp'(w) = -\frac{2}{w^3} + \frac{g_2}{10}w + \frac{g_3}{7}w^3 + \hdots, \quad w \in \mathbb{C},
\end{equation}
and 
\begin{equation}
    \oldwp''(w) = \frac{6}{w^4} + \frac{g_2}{10} + \frac{3 g_3}{7}w^2 + \hdots, \quad w \in \mathbb{C}. 
\end{equation}
We obtain the complete $\dot{a}(t)$ solution given by
\begin{equation}
    \dot{a}(t) = \frac{A}{2} \frac{\oldwp'(t) -E\oldwp''(t)}{\left(\oldwp(t)+\frac{\Omega_k}{12}\right)^2 -C} - \frac{A (\oldwp(t) + \frac{\Omega_k}{12})\oldwp'(t) \left(\oldwp(t) -E\oldwp'(t) +\frac{\Omega_k}{12} \right)}{\left[\left(\oldwp(t)+\frac{\Omega_k}{12}\right)^2 -C\right]^2},
\end{equation}
and, therefore
\begin{eqnarray}
    H(t) = \frac{2\left[\left(\oldwp(t) -\frac{\Omega_k}{12}\right)^2-C\right] \left(- \frac{A (\oldwp(t) + \frac{\Omega_k}{12})\oldwp'(t) \left(\oldwp(t) -E\oldwp'(t) +\frac{\Omega_k}{12} \right)}{\left[\left(\oldwp(t)+\frac{\Omega_k}{12}\right)^2 -C\right]^2} + \frac{A}{2} \frac{\oldwp'(t) -E\oldwp''(t)}{\left(\oldwp(t)+\frac{\Omega_k}{12}\right)^2 -C} \right) }{A\left( \oldwp(t) -E\oldwp'(t) +\frac{\Omega_k}{12} \right)}. ~~
\end{eqnarray}
Finally, we have another indirect reconstruction of $H(t)$ coming from a complete analytic solution for the scale factor. 

\section{Analysis of the Hubble parameter reconstruction}
\label{sec:data-analysis}

In order to quantify the behaviour of the solutions, we can start by rewriting all our piecewise solutions in term of $z$. Using the relation obtained in \cite{CosmicTime} from Lorentz transformations we can derive $H(t)$ relations into $H(z)$. Therefore,
\begin{equation}
    H_0t = \frac{2}{1 + (1+z)^2},
\label{eq:ttoz}
\end{equation}
where the variety of solutions for $H(t)$ are:
\begin{itemize}
\item Direct expression for $H(z)$ is given by: 
\begin{equation}
    H_d(z) = H_0 \sqrt{\Omega_{M,0}(1+z)^3 + \Omega_\Lambda},
\label{eq:fin:direct}
\end{equation}
for late times.
\item For the piecewise solution: 
\begin{equation}
H_{pw} (z) = \left\{ \begin{matrix}
\frac{H_0(1 + (1+z)^2)}{4} & \quad & \tilde{z}_{s} \leq z, \\
\frac{H_0(1 + (1+z)^2)}{3} & \quad & \tilde{z}_{M\Lambda} \leq z, \leq \tilde{z}_{MR} \\
\sqrt{\Omega_\Lambda} H_0 & \quad & z < \tilde{z}_{M\Lambda}.
\end{matrix} 
\right.
\label{eq:Hzpiecewise}
\end{equation}
\item For the mixed fluids as denoted in \cite{galanti_2021_precision} we have: 
\begin{equation}
H_g(z) = \left\{ \begin{matrix} 
\frac{2 \dot{X}(z)}{3 \sqrt{1 -X^2(z)}} \frac{\sin\left[ \frac{1}{3}\arccos(X(z))  \right]}{1 - 2\sin{\left[\frac13\arcsin{\left(X(z)\right)} \right]}}  \quad z > z_* \\
\frac{2 \dot{X}(t)}{3 \sqrt{1 -X^2(t)}} \frac{\cos\left[ \frac{1}{3}\arcsin(X(t))  \right]}{1 + 2\cos{\left[\frac13\arccos{\left(X(t)\right)} \right]}} \quad z_s < z \leq z_* \\
\Omega_{\Lambda}^{1/2}H_0 \frac{A(z;\Omega_\Lambda,H_0)}{B(z;\Omega_\Lambda,H_0)} \quad z < z_s;
\end{matrix} \right.
\label{eq:hubble_analyticz}
\end{equation}
where we wrote $X(t(z)) = X(z)$. Explicitly, these are:
\begin{equation}
    X(z) = 1 -3\frac{\Omega_{M,0}^2}{\Omega_{R,0}^{3/2}}\left( \frac{2}{1+(1+z)^2} \right) + \frac{9}{8}\frac{\Omega_{M,0}^4}{\Omega_{R,0}^3}\left( \frac{2}{1+(1+z)^2} \right)^2,
\end{equation}
and 
\begin{equation}
    \dot{X}(z) = -3\frac{\Omega_{M,0}^2}{\Omega_{R,0}^{3/2}}H_0 + \frac{9}{8}\frac{\Omega_{M,0}^4}{\Omega_{R,0}^3} H_0 \left( \frac{2}{1+(1+z)^2} \right),
\end{equation}
using also: 
\begin{equation}
\begin{split}
    A(z;\Omega_\Lambda,H_0) = a_s^{3/2} \sinh{\left[\frac{3\Omega_\Lambda^{-1/2}}{1+(1+z)^2} -\frac32 \Omega_\Lambda^{1/2} H_0t_s \right]} \\+\left(a_s^3 + \frac{\Omega_{M,0}}{\Omega_\Lambda} \right)^{1/2} \cosh{\left[\frac{3\Omega_\Lambda^{-1/2}}{1+(1+z)^2} -\frac32 \Omega_\Lambda^{1/2} H_0t_s \right]},
\end{split}
\end{equation}
and
\begin{equation}
\begin{split}
    B(z;\Omega_\Lambda,H_0) = a_s^{3/2} \cosh{\left[\frac{3\Omega_\Lambda^{-1/2}}{1+(1+z)^2} -\frac32 \Omega_\Lambda^{1/2} H_0t_s \right]} \\+ \left(a_s^3 + \frac{\Omega_{M,0}}{\Omega_\Lambda} \right)^{1/2} \sinh{\left[\frac{3\Omega_\Lambda^{-1/2}}{1+(1+z)^2} -\frac32 \Omega_\Lambda^{1/2} H_0t_s \right]}.
    \end{split}
\end{equation}

\item For our solutions: 
\begin{equation}
    H(z) = \frac{\dot{a}}{a} =\left\{ \begin{matrix} 
    \frac{H_0\left(\sqrt{4(\Omega_k -\beta)D(z) + (\Omega_{M,0}+2\beta D(z))^2} +2\beta D(z) \right)}{2D(z)\sqrt{4(\Omega_k - \beta)D(z) + (\Omega_{M,0} +2\beta D^2(z))}} \quad a_{RM} < a < a_0, \\
    \sqrt{\Omega_\Lambda}H_0\coth{\left(F+G\right)}, \quad a > a_0,
    \end{matrix} \right.
\label{eq:fin:our}
\end{equation}
where in this case $\beta,D(t),F,G$ are defined in the previous sections. For the transformation $D(t(z)) \to D(z)$, the time-dependant functions can be expressed as
\begin{eqnarray}
    D(z) &=& \left(\frac{2}{1+(1+z)^2} -H_0t_{RM}\right) \nonumber\\ &&
    +  \left( 1+ \frac{\Omega_{M,0}}{2(\Omega_k a_{RM} +\Omega_{M,0})} \right)a_{RM} \sqrt{\Omega_k + \frac{\Omega_{M,0}}{a_{RM}}},
\end{eqnarray}

and for $F$:
\begin{equation}
    F(z) = \sqrt{\Omega_\Lambda}\frac{2}{1+(1+z)} -H_0t_0.
\end{equation}
\item The complete analytical solution via Weierstrass function can be derived by rewriting $\oldwp(t) \to \oldwp(z)$ and using the expression (\ref{eq:ttoz}) to obtain: 
\begin{equation}
    \oldwp(z) = \frac{H_0^2 (1 + (1+z)^2)^2}{4} + \frac{g_2}{5H_0^2(1+(1+z)^2)^2} + \frac{3g_3}{4 H_0^4(1+(1+z)^2)^4}+ \hdots
\end{equation}
where
\begin{equation}
    \oldwp'(z) = H_0^3 (1+(1+z)^2)^3 + \frac{g_2}{5H_0(1+(1+z)^2)} + \frac{8g_3}{7H_0^3(1+(1+z)^2)^3} + \hdots
\end{equation}
and
\begin{equation}
    \oldwp''(z) = \frac{3H_0^4(1+(1+z)^2)^4}{8} + \frac{g_2}{10} + \frac{12 g_3}{7H_0^2(1+(1+z)^2)^2} + \hdots
\end{equation}
\end{itemize}

In a late time scenario, meaning $z \to 0$ and in this regime $\Omega_k \to 0$ and $\Omega_{R,0} \to 0$, the direct is the reduction of the $H(z)$ function to the simple case. When $\Omega_{\text{tot}} = \Omega_{M,0} + \Omega_\Lambda$, then $D=E=B=C=0$. The only term different from zero is $A= \frac{\Omega_{M,0}}{2}$, then $g_2=0$ and $g_3=-\frac{A^2\Omega_\Lambda}{12}$. Therefore,
\begin{equation}
    H(z) = \frac{\oldwp'(z)}{\oldwp(z)},
\end{equation}
and using the previous relations for $\oldwp(z)$, $\oldwp'(z)$ and $\oldwp''(z)$ we can obtain
\begin{equation}
    H_{\oldwp}(z) = \frac{H_0\Omega_\Lambda}{2}(1+(1+z)^2) + \frac{(1-\Omega_\Lambda)^2\Omega_\Lambda}{56} \frac{1}{H_0^2(1+(1+z)^2)^2}.
\label{eq:final:wp}
\end{equation}
We can study these solutions by comparing them with Cosmic Chronometers data obtained from \cite{2018CosmicChronometers}, and considering $\Omega_\Lambda=0.7$ and $H_0=68$, both values at $2\sigma$ according to Planck data \cite{Planck:2018vyg}. Additionally, we consider $\Omega_k = 0.001$ for our solution.

\begin{figure}[H]
    \centering
    \includegraphics[scale=0.45]{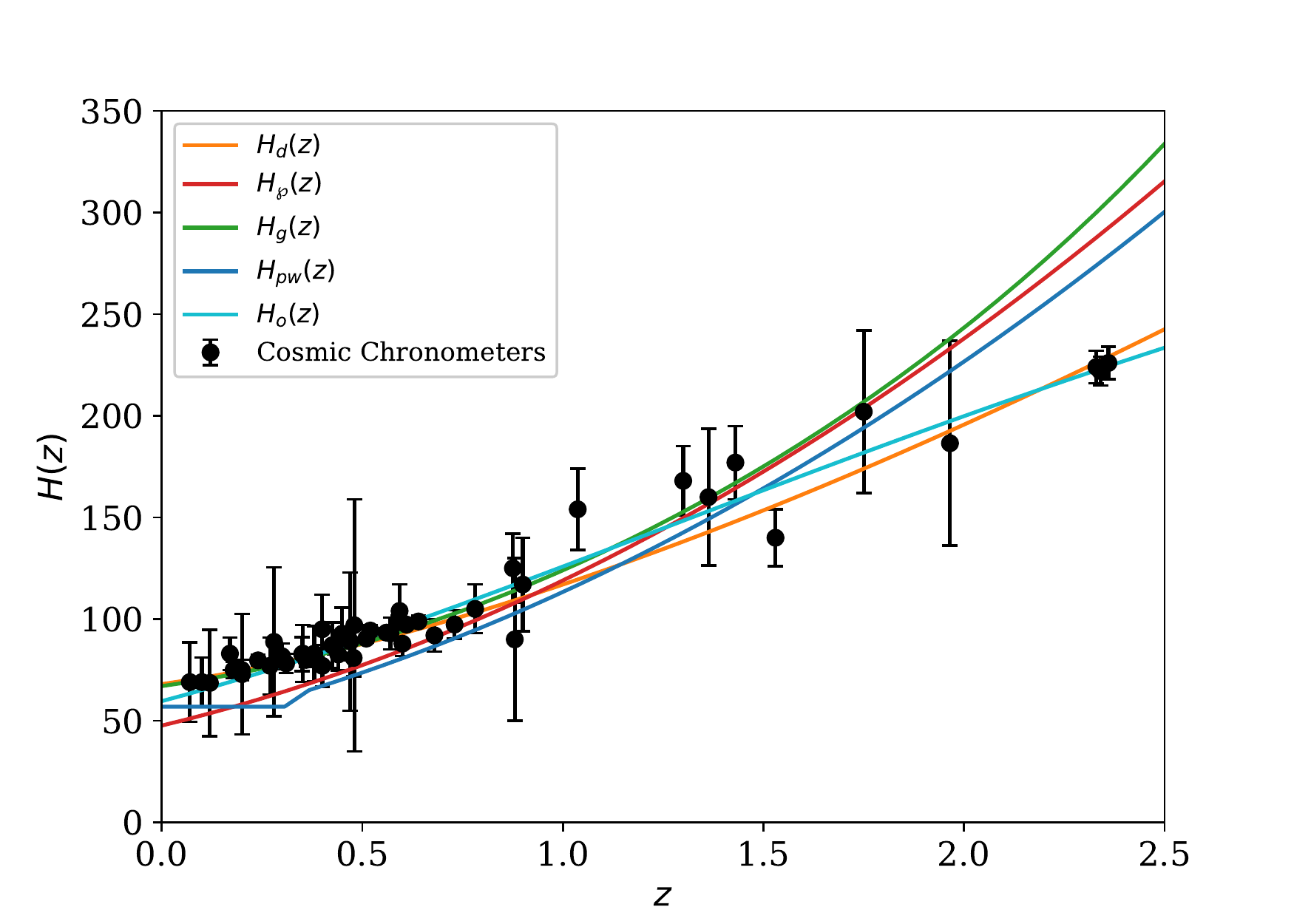}
    \caption{Evolution of the Hubble reconstructions with Planck priors. The light blue line denotes the solution obtained in this work with mixed fluids. The orange line denotes the direct expression. Galanti-like reconstructions are represented by the green line and the piecewise solution by the dark blue curve. Finally, the Weierstrass solution is denoted by the red line. The black dots represents the cosmic chronometers data from \cite{2018CosmicChronometers}.}
    \label{fig:reconstructions}
\end{figure}

\begin{figure}[H]
    \centering
    \includegraphics[scale=0.53]{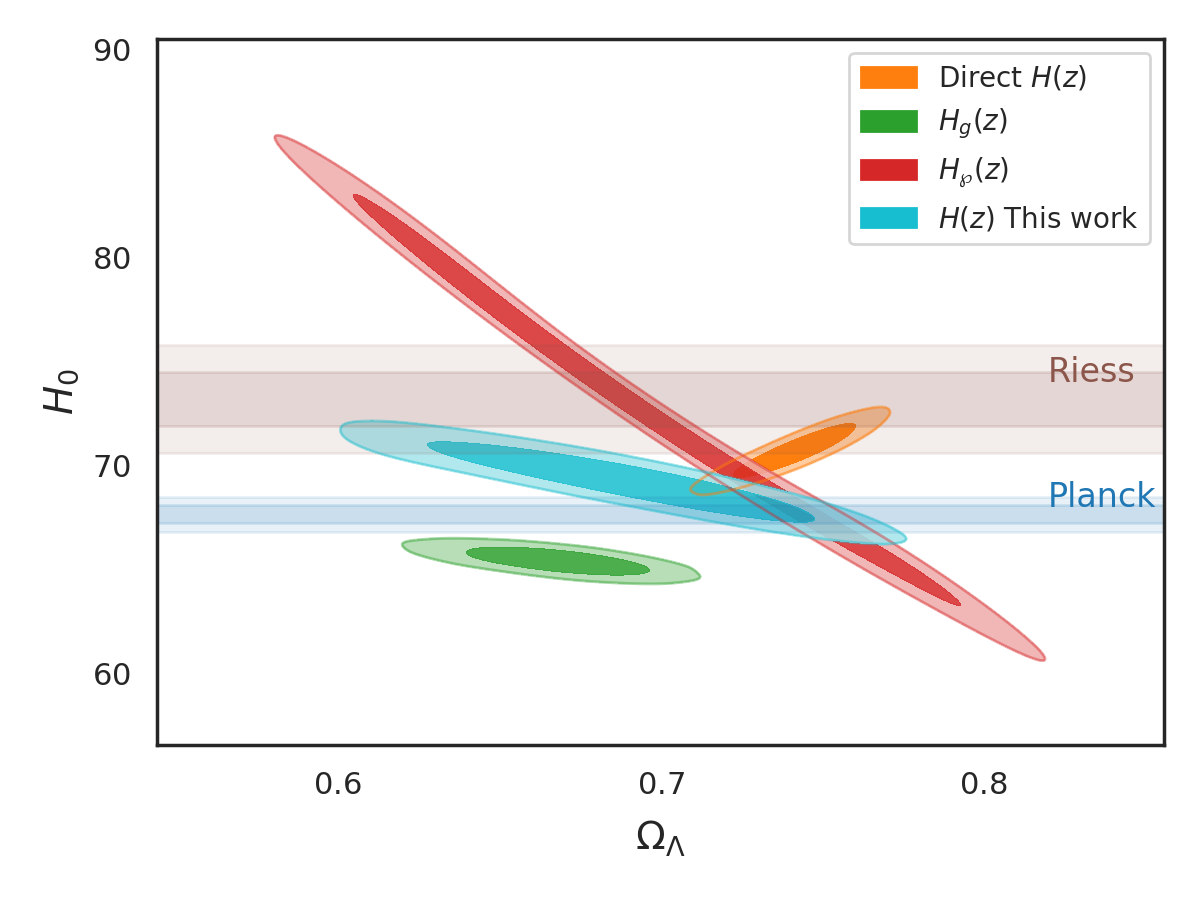}
    \caption{Contour plots for the piecewise solutions analysed. The orange C.L denotes the direct $H(z)$ solution \eqref{eq:fin:direct}, the green C.L denotes the Galanti-Roncadelli solution \eqref{eq:hubble_analyticz}, the red C.L denotes the $\oldwp$-Weierstrass solution \eqref{eq:final:wp} and the blue C.L. denotes our solution \eqref{eq:fin:our}. 
    The brown and blue colors horizontal bands represent the Riess et al. \cite{2021S8Tension} and Planck collaboration \cite{Planck:2018vyg} confidence regions of such $H_0$ priors. }
    \label{fig:my_label}
\end{figure}

Constraining these $H(z)$ expressions with the data indicated, we can obtain the best fit values for $\Omega_\Lambda$ and $H_0$. For that purpose we use a modified version of Emcee library and consider flat priors. The results are reported in Table \ref{tab:Results} and Figure \ref{fig:my_label}.

\begin{table}[h]
    \centering
    \begin{tabular}{|c|cc|c|}
    \hline
        Reconstruction & $\Omega_\Lambda \pm 2\sigma$ & $H_0 \pm 2\sigma$ & $\chi^2$ \\
    \hline
        Direct \eqref{eq:fin:direct} $H_d(z)$ & $0.74^{+0.02}_{-0.03}$ & $70.7^{+1.7}_{-1.7}$ & 27.46 \\
        \emph{Piecewise} \eqref{eq:Hzpiecewise} $H_{pw}(z)$ & - & $75.5^{+1.1}_{-1.1}$ & 556.79 \\
        \emph{Galanti-Roncadelli} \eqref{eq:hubble_analyticz} $H_g(z)$ & $0.69^{+0.04}_{-0.04}$ & $65.41^{+0.88}_{-0.87}$ & 309.21 \\
        This work \eqref{eq:fin:our} $H_o(z)$& $ 0.69^{+0.07}_{-0.08}$ & $69.30^{+2.30}_{-2.41}$ & 30.51 \\
        $\oldwp$-Weierstrass \eqref{eq:final:wp} $H_{\oldwp}(z)$ & $0.70^{+0.1}_{-0.1}$ &  $73.01^{+10}_{-10}$ & 556.98 \\
    \hline
    \end{tabular}
    \caption{. Mean values for the constrained parameters. All confidence intervals are at $2\sigma$.}
    \label{tab:Results}
\end{table}

Notice that in comparison to approach following in \cite{Bochicchio:2011wx}, our solutions are consistent by shells, e.g. from Eq.(\ref{eq:final:wp}), we can classify our function $\oldwp(z)$ by radiation domination, matter domination and dark energy domination solutions in terms of the Hubble flow $H(z)$. At late times, it is possible to recover the simple case, where  $\Omega_{\text{tot}} = \Omega_{M,0} + \Omega_\Lambda$, and the constants (\ref{eq:notations}-\ref{eq:notations2}) evolve according to the late time dynamics.


\section{Conclusions}
\label{sec:conclusions}

The analytical solution obtained from Friedmann equations is useful and well behaved at late times. However, at early times the description of these set of equations are restricted by analysing in a piecewise manner each matter-radiation domination epoch. 
It is worth noticing that the method described here can be, on one hand, applied any epoch that cosmological dynamical system can be reproduce along the Weierstrass functions and an effective cosmological constant plus a curvature term can be defined. As for example, it could be possible to apply this approach to cosmological models that can include scalar fields that mimic the dark sector.
On this path, we have to made assumptions and remove $\Omega_i$-components from the constraint equation to obtain a convergence numerical integration. From Figure \ref{fig:residual} it is clear that analytical solutions work fine for late cosmic epochs, where matter and dark energy are dominant. Current issues at early stage where radiation dominates results in a divergency that does not resemble the numerical integration per se. Therefore, in this work we analyse the solution proposed in \cite{Steiner}, which works efficiently, but the mathematical complexity might be an obstacle for a deeper analysis using current cosmological data. In this line of thought, we derived three analytical solutions: a piecewise-like \eqref{eq:Hzpiecewise}, a $\oldwp$-Weierstrass \eqref{eq:final:wp}, and our solution \eqref{eq:fin:our}. In Figure \ref{fig:reconstructions} we notice that these solution are well behaved at lower redshift for the cosmic chronometer data. Their constrained analysis imply that 
with a precise piecewise methods it is possible to relax $H_0$ tension (see Figure \ref{fig:my_label}), where a Weierstrass approach shows a 1-$\sigma$ agreement with a late time $H_0$ prior from Riess et al, while the direct numerical solution is at 2-$\sigma$ of C.L. 
Our main goal through this Weierstrass approach was to alleviate the Hubble tension, which in these equivalence cosmic epoch a cosmological solution remains with a numerical character, which brings overestimation in the values of its cosmological parameters when constrain them with observations, e.g. when obtaining a best fit value for $H_0$ using local data on a numerical solution that contains radiation and curvature effects, when at this redshift range for this kind of observable these terms are negligible. The possibility to recast through a Weierstrass approach the dynamics of a sector and then apply a reasonable data sample in the redshift range where an exact Weierstrass solution exist, can alleviate this overestimation issue on the parameters, including $H_0$. Of course, we will require better data at higher redshift in order to extend our piecewise approach and tested them, e.g. in cosmological scenarios that could come from alternative theories of gravity.
A further analysis of these approaches employing data at higher redshift will be reported elsewhere.


\backmatter

\bmhead*{Acknowledgments}
CE-R acknowledges the Royal Astronomical Society as FRAS 10147. CE-R and RS-O are supported by PAPIIT UNAM Project TA100122. This work is part of the Cosmostatistics National Group (\href{https://www.nucleares.unam.mx/CosmoNag/index.html}{CosmoNag}) project.

\bmhead*{Data Availability}
All data generated or analysed during this study are included in these published articles:  \cite{Planck:2018vyg} and \cite{2018CosmicChronometers}.

\bibliography{ref}


\end{document}